\documentstyle[aps]{revtex}
\begin{document}
\draft

\title{Edge states of integral quantum Hall states versus \\
edge states of antiferromagnetic quantum spin chains}

\author{Yong Baek Kim}
\address{AT\&T Bell Laboratories, Murray Hill, NJ 07974}

\date{January 11, 1995}

\maketitle

\begin{abstract}

Using the network model representation, it is
shown that the edge states of finite-size integral 
quantum Hall liquid can be regarded as the edge 
states of an SU$(2N)$ open antiferromagnetic quantum 
spin chain in the $N \rightarrow 0$ limit.
The structures of edge states 
in both cases of integer quantum Hall
liquid and an SU$(2N)$ antiferromagnetic quantum spin chain
are compared and the relations between them are 
pointed out. 
This correspondence is used to give qualitative arguments
in favor of the recent results on two-dimensional electron
systems coupled in layers with a large perpendicular
magnetic field.   
In particular, it is shown that the absence
of the localization in the two-dimensional chiral 
surface state of the integral quantum Hall liquid
in the case of the finite-size coupled system can be 
explained by the absence of the gap in the excitation 
spectrum of an SU$(2N)$ ferromagnetic quantum 
spin chain in the $N \rightarrow 0$ limit. 

\end{abstract}

Edge state in finite-size quantum Hall liquids has been 
a subject of interest for many years due to the 
fact that this edge state reflects the topological 
properties of the bulk quantum Hall 
liquids\cite{halperin,wen,macdonald,fisher}. 
On the other hand, recent experiments reveal 
edge states with $S = 1/2$ in the $S = 1$ ($S$ is the 
spin quantum number) antiferromagnetic spin 
chain\cite{hagiwara,glarum}.
The topological character of these edge states in quantum
antiferromagnetic spin chains was recently examined by 
T.~K.~Ng\cite{ng}. 
It is worthwhile to note that, in both systems,
the only gapless excitations are these edge states and
the excitation spectrum of the bulk is generally gapped.
Since both of the quantum Hall liquids and the 
antiferromagnetic quantum spin chains can be 
described by non-linear sigma models (NL$\sigma$M) 
with a topological term\cite{pruisken,haldane,affleck1,affleck2}, 
one may wonder whether the edge states in both systems 
have any similarity.

In this paper, we show that the network model 
representation\cite{chalker0} of the finite-size 
quantum Hall liquid
leads to an SU$(2N)$ open antiferromagnetic quantum 
spin chain in the $N \rightarrow 0$ limit.
In particular, the edge states of the finite quantum Hall
liquids can be considered as the edge states of the
corresponding open spin chain.
Using the results of the large $N$ limit of the spin chain 
and assuming that some of the large $N$ physics survive
in the small $N$ limit, it is shown that the behaviors
of the edge states of the integral quantum Hall liquids
can be understood from the spin chain mapping.

As an application of these results, we apply the mapping 
to the two-dimensional electron systems coupled in layers
with a large perpendicular magnetic field\cite{chalker,balents}.
It is found that the qualitative features of the phase 
diagram of the coupled system can be obtained 
from the ferromagnetically coupled SU$(2N)$ antiferromagnetic
quantum spin chains in the $N \rightarrow 0$ limit. 
For the finite-size quantum Hall systems coupled in layers,
the chiral surface state made of coupled edge states of
each quantum Hall liquid can be regarded as an SU$(2N)$
ferromagnetic quantum spin chain in the 
$N \rightarrow 0$ limit.
It is pointed out that the absence of the localization
in the chiral surface state can be understood in terms 
of the absence of the energy gap in the corresponding
spin chain model.

Recently, for the purpose of studying the plateau 
transitions, D.~H.~Lee and coworkers\cite{dhlee1,wang,dhlee2,dhlee3} 
elaborated the mapping\cite{affleck1,affleck2} between integral 
quantum Hall states and SU$(2N)$ antiferromagnetic 
quantum spin chains in
the $N \rightarrow 0$ limit, where $N$ is the replica index.
The idea is to use the network model of edge states in the 
bulk to describe the quantum Hall phase transitions, 
which takes the advantage of the fact that, 
if an extended state exists, it should be a 
percolating edge state.
The network model of edge states can be 
represented by a collection of chiral fermions
moving in the $y$ direction in an alternating fasion,
{\it e.g.}, $+y$ direction for $x=$ even and 
$-y$ direction for $x=$ odd (see Fig.1)\cite{dhlee1}.
The random tunneling between those chiral
fermions at $(x,y) = (n_x, n_y)$, where $n_x$ and $n_y$ are 
taken as integers, is incorporated by introducing the random
tunneling matrix elements $t(x,y)$.
Here $(n_x, n_y)$s correspond to the 
saddle points of the original network model. 
This network model can be written as\cite{dhlee1}
\begin{eqnarray}
H_{\rm net} &=& \sum_x (-1)^x \int dy \ \psi^{\dagger} (x,y)
{\partial_y \over i} \psi (x,y) \cr
&&- \sum_x \int dy 
\left [ t(x,y) \ \psi^{\dagger} (x+1,y) \psi (x,y) +
{\rm h.c.} \right ] \ ,
\label{originalH}
\end{eqnarray}
where $\psi$ is an annihilation operator of the fermion.
We choose $t(x,y)$ to be random complex numbers with
Gaussian distributions. The random phases of them will
give rise to interference effects which may lead to the 
localization. 

If one takes the $y$ axis as the imaginary time $\tau$ axis,
then the trajectories of the chiral fermions can
be considered as world lines of $(1+1)$ dimensional 
chiral fermions. 
Considering the so-called transport action and taking
the random average of the tunneling matrix elements 
at various space-time points $(x,\tau)$, one can get
an action which describes interacting $2N$ species of 
$(1+1)$ dimensional fermions $\psi_a$ ($a = 1,...,2N$), 
where $N$ is the replica index. 
The factor $2$ comes from the retarded and advanced sectors.
Using the SU$(2N)$ generators 
${\hat S}^b_a = \psi^{\dagger}_a \psi_b - 
\delta_{ab} {1 \over 2N} \sum_c \psi^{\dagger}_c \psi_c$,
where $\psi_a$ is the fermion operator and $a = 1,...,2N$,
the model can be represented by an SU$(2N)$ 
antiferromagnetic quantum spin chain\cite{dhlee1,wang}:
\begin{equation}
H = \sum_x J_x {\rm Tr} \left [ {\hat S}(x + 1){\hat S}(x) \right ] \ .
\label{spinchain}
\end{equation}
where ${\rm Tr}[{\hat A}{\hat B}] = \sum_{ab} A^b_a B^a_b$
and $J_x \sim \langle |t|^2 \rangle (x)$ with 
$\langle \cdots \rangle$ being the random average.
Here it can be shown that $J_{x+2} = J_x$ is satisfied
in general.
Since $H$ commutes with the local density $n(x) = \sum_a
\psi^{\dagger}_a (x) \psi_a (x)$, Hilbert
spaces corresponding to different $\{ n(x) \}$ decouple.
Thus, the ground state of each chain belongs to the Hilbert space
where $n(x) = N$ for all $x$.
In this Hilbert space, a particular representation for SU$(2N)$ is
realized, characterized by a Young tableau with a single column of
length $N$.

In the semiclassical limit, the SU$(2N)$ spin chain of Eq.~\ref{spinchain} 
becomes the ${\rm U}(2N) / {\rm U}(N) \times {\rm U}(N)$ sigma model.
Here the semiclassical limit corresponds to the large representation
limit, {\it i.e.}, a Young tableau with large $M$ ($M$ is an odd integer) 
number of columns of length $N$.
This  ${\rm U}(2N) / {\rm U}(N) \times {\rm U}(N)$ sigma model 
can be written as\cite{dhlee1}
\begin{equation}
{\cal L} = {M \over 16} \sqrt{1 - R^2} {\rm Tr} (\partial_{\mu} Q)^2
+ {M \over 16} (1 - R) \epsilon_{\mu \nu} {\rm Tr} \left [
Q \partial_{\mu} Q \partial_{\nu} Q \right ] \ ,
\label{sigma}
\end{equation}
where $\mu = x,\tau$, $Q = u^{\dagger} \Lambda u$, and
$u$ is a $2N \times 2N$ unitary matrix.
$\Lambda$ is a diagonal matrix with $\Lambda^a_a = 1$ for $a \le N$
and $\Lambda^a_a = -1$ for $N < a \le 2N$.
Here $R = [J_{x+1} - J_{x}]/[J_{x+1} + J_{x}]$ measures the degree
of dimerization.
The second term is the so-called topological term.
For $M = {\rm odd}$, the sigma model is massless at $R = 0$ due to 
the presence of the topological term while
it is massive when $R \not= 0$.
Thus, we can identify the transition from $R < 0$ to $R > 0$ as
the plateau transition because the change of the single particle
energy corresponds to the change in the degree of 
dimerization $R$\cite{dhlee1}.
Note that this transition is nothing but the transition from 
one dimerized state ($R < 0$) to the other dimerized state 
($R > 0$) in the corresponding quantum spin chain and these dimerized
states correspond to the integral quantum Hall states in the original 
problem.
At the critical point ($R = 0$), by comparing this sigma model with
the Pruisken's sigma model\cite{pruisken}, it can be noticed that 
$\sigma_{xy} = M/2$.

Let us consider finite-size quantum Hall liquid which has boundaries
at $x = 0, L$. 
The corresponding network model can be written as
\begin{eqnarray}
H_{\rm net} &=& \sum_{0 \le x \le L} (-1)^x \int dy \
\psi^{\dagger} (x,y) {\partial_y \over i} \psi (x,y) \cr
&&- \sum_{0 \le x \le L} \int dy 
\left [ t(x,y) \ \psi^{\dagger} (x+1,y) \psi (x,y) +
{\rm h.c.} \right ] \ ,
\end{eqnarray} 
where $L$ is taken as an odd integer so that the directions
of the edge states at $x = 0$ and $x = L$ are opposite.
Using the same procedure employed for the derivation of Eq.~\ref{spinchain},
the model can be represented by the following open SU$(2N)$ 
quantum antiferromagnetic spin chain:
\begin{equation}
H = \sum_{0 \le x \le L} 
J_x {\rm Tr} \left [ {\hat S}(x + 1){\hat S}(x) \right ] \ .
\end{equation}
Note that this spin chain has an even number, $L+1$, of sites.
The appearance of the additional edge state in this finite
spin chain as the degree of dimerization $R$ changes the
sign can be understood as follows.
Suppose that we start with a dimerized spin chain where
$J_x \not= 0$ for $x = {\rm even}$ and $J_x = 0$ for
$x = {\rm odd}$. In this case, there is no edge state.
Now we change the degree of dimerization such that 
$J_x = 0$ for $x = {\rm even}$ and $J_x \not= 0$ for 
$x = {\rm odd}$. In this case, two free spins will be left
at $x = 0, L$ and they are completely decoupled from 
the bulk of the spin chain.
When $R$ is small but finite, the edge states are not
completely decoupled and they decay into the spin chain
with a length scale given by the coherence length $\xi$
which is inversely proportional to the gap of the 
dimerized spin chain\cite{ng}.
In the language of the original quantum Hall liquid,
additional edge state appears as the interger quantum Hall
transition occurs ($R = 0$ at the transition in the spin 
chain) and it decays into the bulk with a length scale
given by the localization length $\xi$ which is
inversely proportional to the gap of the quantum Hall liquid.       

Recently, T.~K.~Ng\cite{ng} used the $CP^1$ representation or a gauge field 
picture of the SU$(2)$ antiferromagnetic quantum spin chain to 
explain the behaviors of edge states of the antiferromagnetic spin 
chains with the spin quantum number $S$. 
We will mainly follow and generalize his analyses to get insights
about the edge states of the SU$(2N)$ quantum spin chains. 
In particular, we will use the results of the large $N$ limit
of an SU$(2N)$ antiferromagnetic spin chain\cite{ng,affleck1,affleck2}.
The hope is that the edge state picture obtained from the
large $N$ limit will survive in the small $N$ limit as far
as there is a gap in the bulk excitation spectrum 
even though it is well known that some of the large $N$ 
results do not apply to the case of $S = 1/2$ and $N = 1$ 
(half-integer spin chain).
This expectation can be supported by comparing the known behaviors 
of the edge states of quantum Hall liquids with the physics coming
out from the large $N$ analysis. 
Thus, it can be also a good way of testing the usefulness of
the large $N$ results to the small $N$ limit.
 
In Ref.~\cite{affleck1} and \cite{sachdev}, 
various representations of the SU$(2N)$ 
antiferromagnetic spin chain were examined.
In particular, at $x = {\rm odd}$, one can use a Young tableau 
with $0 < n < 2N$ rows and $M$ columns.
On the other hand, at $x = {\rm even}$, a Young tableau with
$2N-n$ rows and $M$ columns can be taken.
For the case of $N=1$ (SU$(2)$), all representations have $n=1$
and $M = 2S$ where $S$ is the spin quantum number.
In the semiclassical limit, it can be shown that the model can
be described by the (1+1) dimensional 
${\rm U}(2N) / {\rm U}(n) \times {\rm U}(2N-n)$ NL$\sigma$M
with a topological term\cite{sachdev}.
In this paper, we will mainly use the results of the large 
$N$ limit of $n=1$ representation even though the previously
described sigma model corresponds to $n = N$.
The advantage is that the model can be mapped to $CP^{2N-1}$
model\cite{dadda,witten} with a topological term and this gauge 
field description
turns out to be useful for the analysis of the edge states.
For general $n$, the corresponding gauge field theories can
be formulated, but the gauge symmetry becomes 
${\rm U}(n) = {\rm U}(1) \times {\rm SU}(n)$\cite{sachdev,macfarlane}. 
We expect that the qualitative structure of the edge state 
obtained from $n=1$ case would be the same as that of $n=N$ case
as far as there is an excitation gap in the bulk
of the SU$(2N)$ spin chain\cite{sachdev}.

The ${\rm U}(2N) / {\rm U}(1) \times {\rm U}(2N-1)$ NL$\sigma$M
with a topological term can be rewritten as the $CP^{2N-1}$
model with a topological term:
\begin{equation}
{\cal L} = {2 \over g}|(\partial_{\mu} + iA_{\mu}) z^{\alpha}|^2
- i{\theta \over 2\pi}\varepsilon^{\mu \nu} \partial_{\mu} A_{\nu} \ ,
\end{equation}
where $z^{\alpha}$ are $2N$ complex fields ($\alpha = 1,...,2N$)
satisfying $\sum_{\alpha} |z^{\alpha}|^2 = 1$, $g \sim 2/M$ 
and $\theta = M \pi$. The second term is called the topological term.
Note that $M = 2S$ when $N = 2$.
In the large $N$ limit, at the gaussian level 
({\it i.e.}, to the lowest order in $1/(2N)$), the theory has
$2N$ massive free bosons with the gap $m \sim e^{-\pi / (2Ng)}$.
In the $1/(2N)$ expansion, an additional term 
${N \over 4 e^2} F^2_{\mu \nu}$ is generated to the $1/(2N)$ order.
As the result, the effective low energy theory in the large
$N$ limit can be written as\cite{sachdev,dadda,witten} 
\begin{equation}
{\cal L}_{\rm eff} = {2 \over g}|(\partial_{\mu} + iA_{\mu}) z^{\alpha}|^2
+ m^2 |z^{\alpha}|^2
- i{\theta \over 2\pi}\varepsilon^{\mu \nu} \partial_{\mu} A_{\nu}
+ {N \over 4 e^2} F^2_{\mu \nu} \ ,
\end{equation}
where $e^2 \sim m^2$ and 
$F_{\mu \nu} = \partial_{\mu} A_{\nu} - \partial_{\nu} A_{\mu}$.
Since U$(1)$ gauge theory in $(1+1)$ dimension has linear
Coulomb force, the $z$ bosons are confining\cite{sachdev,dadda,witten}.
It should be noted that, even though the physics of the large $N$ 
limit is believed to be qualitatively correct for most of the 
values of $\theta$, it is well known that there is no gap
in the excitation spectrum for $N=2$ if $\theta = (2k+1) \pi$ or 
$M = 2k+1$ (half-integer spin chain) while the large $N$ theory
predicts a gap.
Therefore, for the behaviors around $M = 2k+1$, one should be careful
in the interpretation for the small $N$ limit which we are 
interested in.

As pointed out in Ref.~\cite{ng}, for the finite SU$(2N)$ spin chains, 
the effects of the topological term are crucial for the 
understanding the behavior of the edge states.
Since $\varepsilon^{\mu \nu} \partial_{\mu} A_{\nu}$ corresponds
to the electric field of the U$(1)$ gauge field, the effect of
the topological term is the presence of a uniform external field
$E_{\rm ext} = {\theta \over 2\pi}e = {M \over 2}e$ across the
one-dimensional universe\cite{coleman}.
In the case of the finite quantum spin chain, the one-dimensioanl 
universe has a finite size and the uniform electric field can be
considered as arising from the external charges 
$\pm {\theta \over 2\pi}e$ at $x = 0, L$\cite{ng}.
As shown by Read and Sachdev\cite{sachdev}, the electric field of the U$(1)$
gauge theory corresponds to the finite dimerization order-parameter
$\langle {\rm Tr}[{\hat S}(x+1){\hat S}(x) - 
{\hat S}(x){\hat S}(x-1)] \rangle$. Therefore, the spin chain is 
generically dimerized due to the finite electric field in the 
corresponding gauge theory, so there is an excitation gap.
For example, when $\theta < \pi$, the spin chain is dimerized. 
However, as $\theta$ becomes $\pi$ ($M = 1$) or larger than $\pi$, 
it is energetically favorable that a pair of $z$ bosons
are nucleated from the vacuum and try to screen the external
charges at $x = 0, L$. In this case, the external charges are 
overscreened and the effective external electric field  becomes
$E_{\rm eff} = \left ( {\theta \over 2 \pi} - 1 \right )e$.
Due to this overscreening, the direction of the electric field
is abruptly reversed as $\theta$ crosses $\pi$ and it implies
that the dimerization order-parameter also changes sign.
The same phenomena occur whenever $\theta$ crosses 
$(2k+1) \pi$ or $M$ crosses $2k +1$.
That is, at each time when $\theta$ crosses $(2k+1) \pi$,
one additional pair of $z$ bosons are nucleated from
the vacuum and go to the opposite edges of the universe.
At the same time, the direction of the electric field and 
the sign of the dimerization order parameter changes. 

Now let us translate these phenamena into the language of the
interger quantum phase transition. Note that $\theta = (2k+1) \pi$
or $M = 2k+1$ corresponds to $\sigma_{xy} = M/2 = (2k+1)/2$.
Also the quantum Hall liquid has a gap if 
$\sigma_{xy} \not= M/2$.    
At each time $\sigma_{xy}$ crosses $M/2$,
an additional edge state is generated and the number of 
edge state is nothing but the total number $z$ bosons
nucleated from the vacuum to screen the external electric field.
Unfortunately, the large $N$ theory predicts first order 
transitions at each $\sigma_{xy} = M/2$ due to the abrupt changes
of the dimerization order parameter.
Since these transitions in quantum Hall states are consistent
with the second order phase transions, one may argue that
the first order transition of the large $N$ theory
does not apply to the small $N$ limit.
In particular, when the degree of dimerization $R$ is small,
$\theta = M \pi (1-R)$ and $J_x = J [1 - R (-1)^x]$, thus
we expect that the localization length $\xi$ diverges as $R$
goes to zero.

The wave function of the $z$ bosons
at the edge has a characteristic size of $\xi \sim m^{-1}$ when the
bulk has an excitation gap $m$.
This implies that the edge states of the quantum Hall liquid
decay into the bulk with a length scale given by the 
localization length. 
Since the wave functions of the $z$ bosons
are smeared out around the edge, the corresponding 
effective electric field is not uniform along the length
scale $\xi$. Correspondingly, in the spin chain, the dimerization is
not uniform around the edge. Note that the spatial dependence 
of the degree of dimerization in the spin chain corresponds 
to the spatial dependence of the filling fraction or the 
spatial dependence of the density of electrons.
Therefore, the presence of the finite-size edge states decaying
into the bulk in the spin chain can be translated into the 
inhomogeneous density distribution of the electrons around 
the edge of the quantum Hall liquid. 
This is nothing but the oscillation of the 
electron density (Friedel oscillation) around the edge of the 
quantum Hall liquid.     
 
As an application of the above analysis, we will consider
coupled layered two-dimensional samples which would form 
individually integral quantum Hall states in a large perpendicular 
magnetic field if there was no interlayer coupling.
This problem was recently studied numerically by Chalker and 
Dohmen\cite{chalker} who used an anisotropic generalization of 
the two-dimensional network model. 
Using the sigma model approach, Balents and Fisher\cite{balents} 
also studied it analytically. 
This problem arises in two dimensional 
multilayer electron system and the Bechgaard salts\cite{balicas}.
The main conclusions of these works are as follows.
When the system is infinite in the plane as well as
in the $z$-direction, it is found that there is a band 
of extended states around the band center when the
interlayer tunneling amplitude is not zero.
In this case, as the single particle energy $E$
is varied, there are successive two transitions which 
correspond to insulator-metal (or quantum-Hall-liquid-metal) 
and metal-quantum-Hall-liquid transitions\cite{chalker}.
For layered materials, when the system is finite in the
plane but it is inifinite in the $z$ direction, there is 
the chiral surface of edge states which are coupled by the 
interlayer tunneling amplitude.
It turns out that there is no localization
in the disordered chiral surface even though all the 
electronic states are localized in the usual two 
dimensional disordered electron system\cite{chalker,balents}.

We are going to use the network model of Chalker and
Dohmen\cite{chalker} (see Fig.2), which can be written in terms of the 
previously introduced chiral fermions.
\begin{eqnarray}
H_{\rm net} &=& \sum_{x,z} (-1)^x \int dy \ \psi^{\dagger} (x,y,z)
{\partial_y \over i} \psi (x,y,z) \cr
&&- \sum_{x,z} \int dy 
\left [ t (x,y,z) \ \psi^{\dagger} (x+1,y,z) \psi (x,y,z) +
{\rm h.c.} \right ] \cr
&&- \sum_{x,z} \int dy 
\left [ t_{\perp} (x,y,z) \ \psi^{\dagger} (x,y,z+1) \psi (x,y,z) +
{\rm h.c.} \right ] \ ,
\end{eqnarray}
where $\psi (x,y,z)$ ($z$ is an integer such that $1 \le z < \infty$) 
is the fermion annihilation operator of the $z$th layer. 
The scattering processes of the
network are now modeled by tunneling ($t$) between neighboring sites with
the same layer index, as well as tunneling ($t_{\perp}$) between
adjacent layers at the same $(x,y)$-site.
We choose $t(x,y,z)$ and $t_{\perp}(x,y,z)$ to be independent random complex
numbers with Gaussian distributions.
Using the same procedure of getting Eq.~\ref{spinchain} from Eq.~\ref{originalH}, 
after the random averages, the model can be mapped to the system of 
coupled SU$(2N)$ quantum spin chains.
When the size of the sample is finite in the $x$ direction such that 
$0 \le x \le L$, the coupled spin chain model can be written as
\begin{eqnarray}
H &=&
 \sum_{0 \le x \le L} \sum_{z} \ J_{x,z}
       \ {\rm Tr} \left [ {\hat S} (x+1,z) {\hat S} (x,z) \right ] \cr &&
+ \sum_{0 \le x \le L}  \sum_{z} \ J_{\perp}
       \ {\rm Tr} \left [ {\hat S} (x,z+1) {\hat S} (x,z) \right ] \ ,
\end{eqnarray}
where $J_{x,z} = \langle |t|^2 \rangle (x,z)$ and 
$J_{\perp} = - \langle |t_{\perp}|^2 \rangle$\cite{kim}.
The above Hamiltonian gives nearest-neighbor antiferromagnetic
couplings ($J_{x,z} > 0$) in each spin chain $z$. On the other hand, for
each $x$, the coupling between spins on adjacent chains, $z+1$ and $z$,
is ferromagnetic ($J_{\perp} < 0$) and uniform.
In terms of the original problem, the energy gap of the coupled layered
system corresponds to the existence of an
energy gap in the spectrum of this spin chain (in the replica limit of
$N \rightarrow 0$).

In the first place, let us consider the infinite sample 
($-\infty < x < \infty$).
The two-layer case ($z = 1,2$) has been studied\cite{wang,dkklee} and 
it is found that there are two transitions as the single particle 
energy is changed,
where the change of the single particle energy corresponds to the 
change in the ratio between two different measures of dimerization
$R_{z} = [J_{x+1,z} - J_{x,z}]/[J_{x+1,z} + J_{x,z}]$ ($z=1,2$). 
It is found that three massive phases are separated by two transitions which
correspond to massless points in the parameter space of the model.
These are two dimerized states and the intervening Haldane phase.
One can expect that, as the number of chains is increased, the change
in single particle energy in the original layered material depends 
on more and more parameters $R_z$
so that there will be many transitions which correspond to gapless points.
Also the energy gaps of the intervening massive phases become smaller
and smaller. 
Therefore, in the limit of infinite number of chains, except two 
dimerized phases in the tails of the band, the intervening 
phase has no energy gap.
These results imply that, in the original problem,
there are two localized phases (insulator or quantum-Hall-liquid)
in the tails of the single particle energy band and there is a 
band of extended states between them, which corresponds to a 
metallic phase.  
Note that this conclusion from the coupled spin chains is 
consistent with the results of the numerical calculation\cite{chalker}.

Now let us come back to the case of the finite-size sample.
Suppose that the system is in the quantum-Hall-liquid phase.
If the single particle energy is far away from those of the 
extended states, the localization length is very short so that
we can consider the chiral surface state being almost 
decoupled from the bulk of the sample.
In this case, for instance, the chiral surface states at 
$x = L$ corresponds to the following spin chain model.
\begin{equation}
H = \sum_z J_{\perp} \ {\rm Tr} \left 
[ {\hat S} (x=L,z+1) {\hat S} (x=L,z) \right ] \ .
\end{equation}
Note that $J_{\perp}$ is ferromagnetic so that the model
becomes a uniform ferromagnetic SU$(2N)$ spin chain.
If one assumes that the results of finite $N$ still apply
to the $N \rightarrow 0$ limit, this model
has no energy gap.
This implies that, in the original problem of the chiral surface
state, there is no localization even though the system is
disordered.

In summary, it is shown that the edge states of integral 
quantum Hall liquid can be mapped to the edge states of
an SU$(2N)$ quantum antiferromagnetic spin chain in the
$N \rightarrow 0$ limit. Using the gauge field theory
of the large $N$ limit of the spin chain and assuming that 
some of the large $N$ results are robust even in the small
$N$ limit, the relation between two edge states is studied.
This relation is used to confirm some of the recent results
on two dimensional non-interacting electron systems coupled 
in layers. In particular, it is shown that the chiral
surface state of the finite-size coupled-layered system
can be regarded as an SU$(2N)$ quantum ferromagnetic spin
chain in the $N \rightarrow 0$ limit and the absence of 
the localization in the disordered
chiral surface state can be explained
by the absence of the excitation gap in the 
corresponding ferromagnetic spin chain.

\acknowledgments
We would like to thank B.~I.~Halperin, D.~K.~K.~Lee, 
P.~B.~Littlewood, A.~J.~Millis, 
T.~-K.~Ng, A.~M.~Sengupta, B.~I.~Shraiman and C.~M.~Varma for 
helpful discussions.

\begin{figure}
\caption{
A representation of the network model for the two dimensional
integral quantum Hall state in terms of 
chiral fermions (thick solid lines) with alternating directions.
The dotted lines represent the tunnelings between chiral
fermions at the saddle points.}
\end{figure}

\begin{figure}
\caption{
A representation of the network model for the
two dimensional electron systems coupled in layers
with a large perpendicular magnetic field.
The dotted and the thin soild lines represent the 
in-plane and the interlayer tunnelings respectively.}
\end{figure}

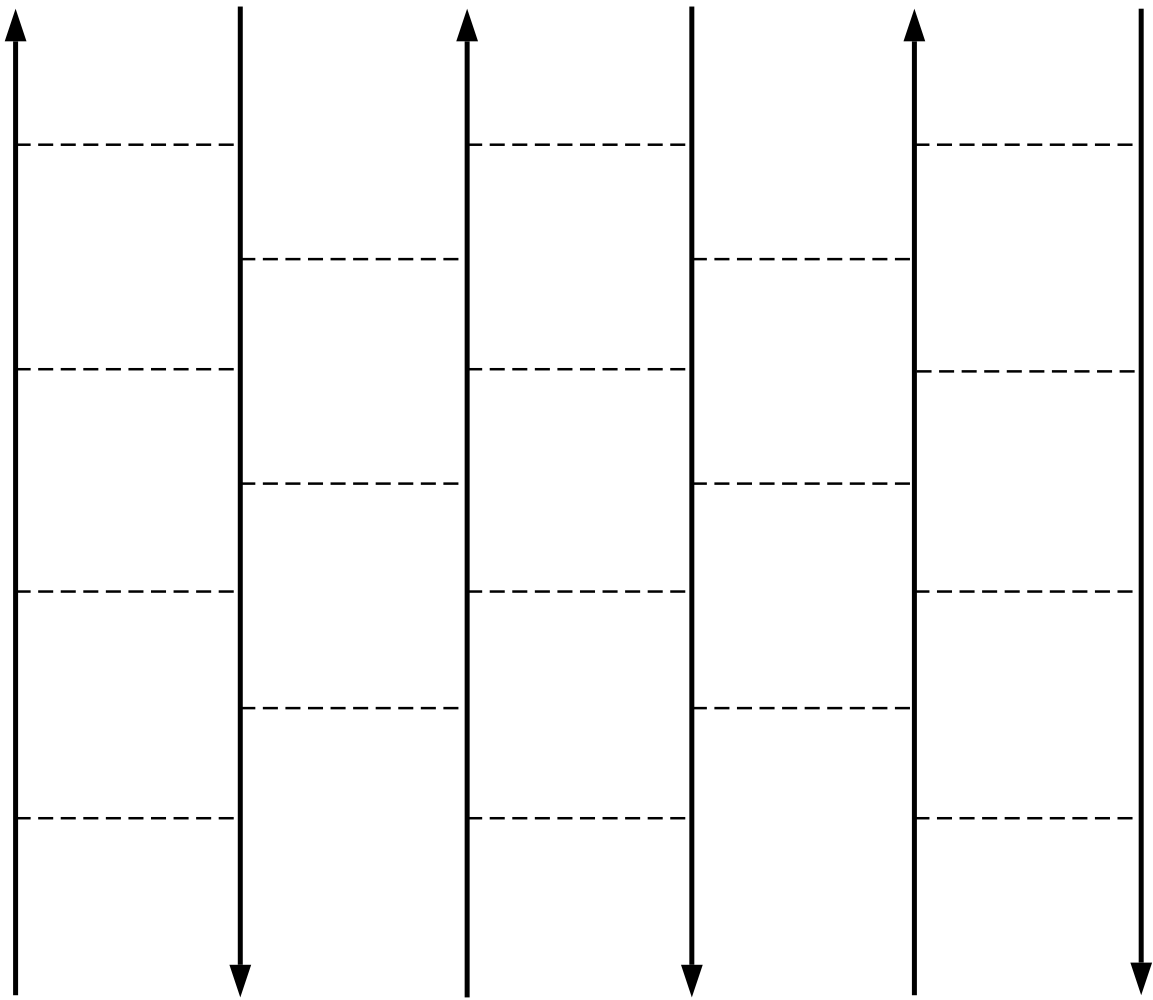
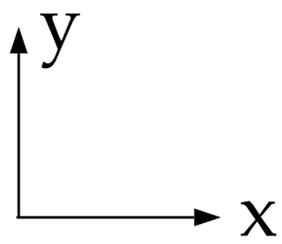
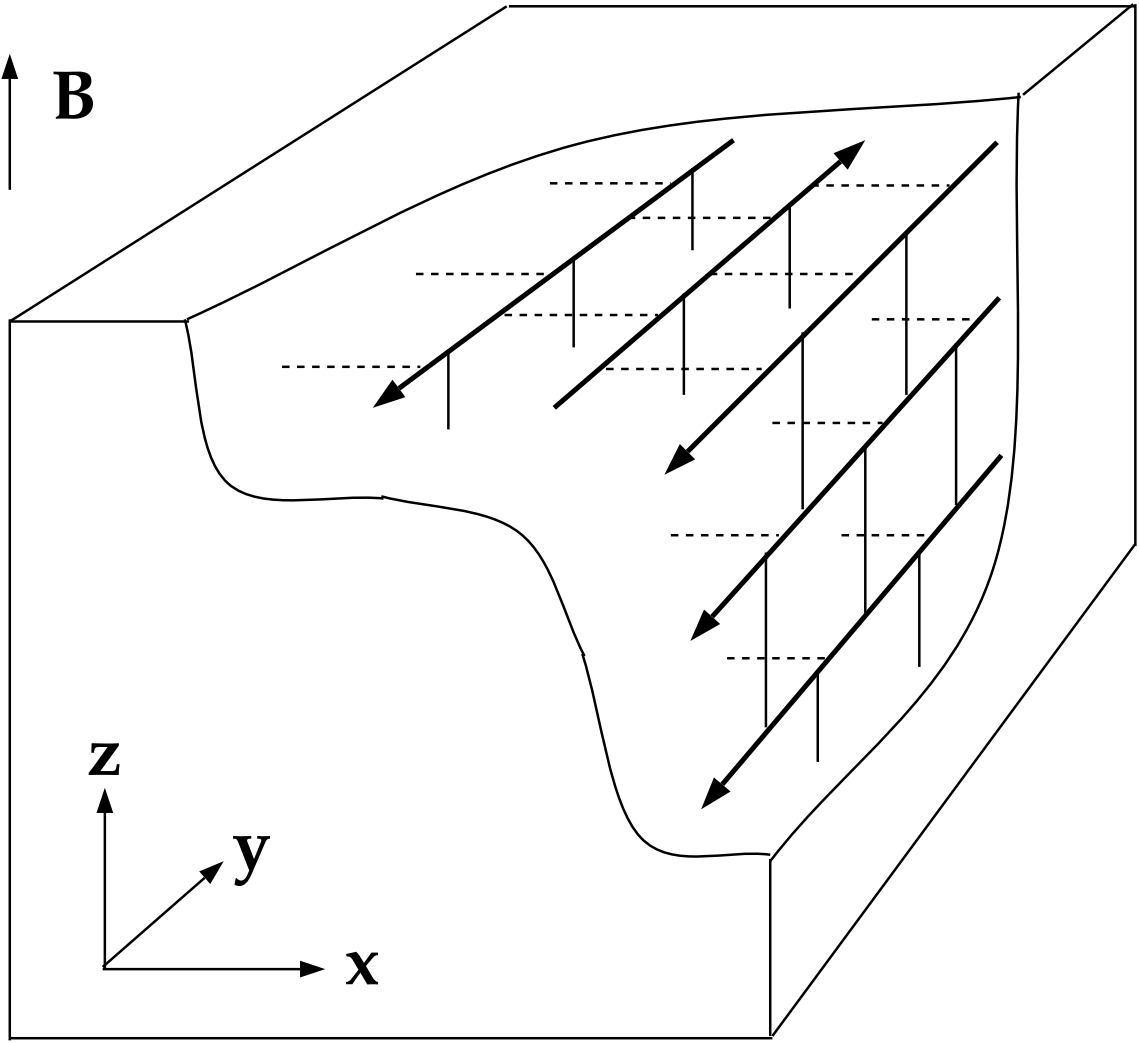
\end{document}